\documentclass[aps,pre,groupedaddress,twocolumn]{revtex4}

\usepackage{hyperref}
\usepackage{subfigure}
\usepackage{color}
\usepackage{graphicx}
\usepackage{natbib}
\usepackage{doi}

\begin{document}
\bibliographystyle{plainnat}

\expandafter\ifx\csname urlprefix\endcsname\relax\def\urlprefix{URL }\fi

\title{\large
Hydrodynamics of the Physical Vacuum:
II.~Vorticity dynamics}

\large
\author{Valeriy I. Sbitnev}
\email{valery.sbitnev@gmail.com}
\address{St. Petersburg B. P. Konstantinov Nuclear Physics Institute, NRC Kurchatov Institute, Gatchina, Leningrad district, 188350, Russia;\\
 Department of Electrical Engineering and Computer Sciences, University of California, Berkeley, Berkeley, CA 94720, USA
}


\date{\today}

\begin{abstract}
  Physical vacuum is a special superfluid medium populated by enormous amount of virtual particle-antiparticle pairs.
   Its motion is described by the modified Navier-Stokes equation: 
 (a)~the pressure gradient divided by the mass density is replaced by the gradient from the quantum potential; 
 (b)~time-averaged the viscosity vanishes, but its variance is not zero. 
 Vortex structures arising in this medium show infinitely long lifetime owing to zero average viscosity. 
 The nonzero variance is conditioned by exchanging the vortex energy with zero-point vacuum fluctuations.
 The vortex has a non-zero core where the orbital speed vanishes. 
 The speed reaches a maximal value on the core wall and further it decreases monotonically.
 The vortex trembles around some average value and possesses by  infinite life time.
 The vortex ball resulting from topological transformation of the vortex ring
  is considered as a model of a particle with spin.
  Anomalous magnetic moment of electron is computed.


\end{abstract}

\maketitle

\section{\label{sec1}Introduction}

 The Navier-Stokes equation possesses by a rich set of solutions describing motions of fluids ranging from simple irrotational forms up to complicated vortex dynamics. Thanks to the minor modification of this equation one may get the equation which describes a motion of a special superfluid medium - the physical vacuum populated by enormous amount of virtual particle-antiparticle pairs. 

 The modified Navier-Stokes equation has been described in the previous article~\cite{Sbitnev20xx}
 and it has the following view
\begin{equation}
 m\biggl(
 {{\partial {\vec {\mathit v}}}\over{\partial\,t}}
 + ({\vec {\mathit v}}\cdot\nabla){\vec {\mathit v}}
       \biggr) 
  =  {{{\vec{\mathit F}}}\over{N}}
   \;-\; \nabla Q
 \; +\; \nu(t)\,\nabla^{\,2}m{\vec {\mathit v}}.
\label{eq==1}
\end{equation} 
 It is accompanied by the continuity equation
\begin{equation}
 {{\partial\,\rho_{_{M}}}\over{\partial\,t}} +(\nabla\cdot{\vec{\mathit v}})\rho_{_{M}} = 0.
\label{eq==2}
\end{equation}
 Here $\rho_{_{M}}=mN/\Delta V=m\rho$ is a mass density of the fluid in the volume $\Delta V$
 populated by $N$ particles, each has the mass $m$.
 The term  $Q$ is the quantum potential and ${\vec{\mathit F}}/N$ is an external force per particle.  
 We believe, this force is conservative.
 The kinetic viscosity of the fluid, $\nu(t)$, is a fluctuating function of the time with the expectation equal to zero.
  
\begin{figure}[htb!]
 \centering
  \begin{picture}(200,70)(-40,5)
      \includegraphics[scale=0.2, angle=0]{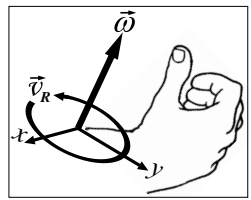}
  \end{picture}
  \caption{
Right hand rule: rotation with the orbital velocity ${\vec{\mathit v}}_{_{R}}$ lies in the plane $(x, y)$.
 The vorticity vector ${\vec\omega}$ is oriented perpendicular to this plane in direction indicated by thumb   }
  \label{fig==1}
\end{figure}

      The velocity ${\vec{\mathit v}}$ is represented consisting of two components~\cite{KunduCohen2002}:
\begin{equation}
\label{eq==3}
  {\vec{\mathit v}} = {\vec{\mathit v}}_{_{S}} + {\vec{\mathit v}}_{_{R}},
\end{equation}
  irrotational, ${\vec{\mathit v}}_{_{S}}$, and solenoidal,  ${\vec{\mathit v}}_{_{R}}$. 
   Here subscripts $S$ and $R$ point to existence of scalar
 and vector (rotatory) potentials  underlying emergence of these  velocities.
  These velocities relate to vortex-free
 and vortex motions of the fluid medium, respectively. 
 Scalar and vector fields underlie of manifestation of these two types of velocities.
 They satisfy the following equations
\begin{equation}
\label{eq==4}
 \left\{
    \matrix{
           (\nabla\cdot{\vec{\mathit v}}_{_{S}}) \ne 0, & [\nabla\times{\vec{\mathit v}}_{_{S}}]=0, \cr
           (\nabla\cdot{\vec{\mathit v}}_{_{R}})  =  0, &\,\, [\nabla\times{\vec{\mathit v}}_{_{R}}]={\vec\omega}. \cr
           }
 \right.
\end{equation}
 Here the vector $\vec\omega$ is called the vorticity. 
 This vector is oriented perpendicularly to the plane of rotation according to the right hand rule, Fig.~\ref{fig==1}.
 
  Here we shall study the vortex dynamics.

    The article is organized by the following manner.
 Sec.~\ref{sec2} gives a general presentation of the vortex dynamics. Sec.~\ref{sec3} describes a topological transformation of the helicoidal vortex ring to the vortex ball (bubble). Here we discuss existence of different flows on surface of the ball. In Sec.~\ref{sec4} we propose the vortex ball as a possible model of the spin 1/2 particle. We compute also the anomalous magnetic moment of electron. Sec.~\ref{sec5} gives concluding remarks.
 
\section{\label{sec2}Vortex dynamics}

  By multiplying Eq.~(\ref{eq==1}) by the operator curl from the left we obtain the equation for the vorticity
\begin{equation}
 {{\partial\, {\vec\omega}}\over{\partial\,t}}
 + ({\vec\omega}\cdot\nabla){\vec{\mathit v}}
 = \nu(\,t\,)\nabla^{2}{\vec\omega}.
\label{eq==5}
\end{equation}
 The rightmost term describes dissipation of the
energy stored in the vortex. As a result, the vortex with the lapse of time should disappear.
By omitting this rightmost term ($\nu=0$) we open the action of the Helmholtz theorem~\cite{Lighthill1986}: (i)~if fluid particles form, in any moment of the time, a vortex line, then the same particles support the vortex line both in the past and in the future; (ii)~ensemble of the vortex lines traced through a closed contour forms a vortex tube; (iii)~intensity of the vortex tube is constant along its length and does not change in time. The vortex tube (a)~either goes to infinity by both endings; (b)~these endings lean on boundary walls containing the fluid; or (c)~these endings are locked to each on other forming a vortex ring.

 Assuming that the fluid is a physical vacuum, which meets the requirements specified in~\cite{Sbitnev20xx},
  we must say that the viscosity vanishes. 
 In that case, the vorticity $\vec\omega$ is concentrated in the center of the vortex, i.e., in the point. Its mathematical representation is $\delta$-function. This singularity can be a source of possible divergences of computations in further.

We shall not remove the viscosity. Instead of that, we hypothesize that the viscosity fluctuates around zero. Its expectation is zero. These fluctuations represent, in fact, exchanging of the stored energy in the vortex with the zero-point vacuum oscillations. 
The vortex does not disappear completely but it may live so long as possible. 
In other words, the vortex is  a  long-lived  object. 

 Here we shall consider a simple model of such a picture. Let us look on the vortex tube in its cross-section which is oriented along the $z$-axis and its center is placed in the coordinate origin of the plane $(x, y)$.
  Eq.~(\ref{eq==5}), written down in the cross-section of the vortex, looks as follows
\begin{equation}
  {{\partial\, {\omega}}\over{\partial\,t}} =
 \nu(\,t\,)\Biggl(
    {{\partial^{\,2}\omega}\over{\partial\,r^{2}}}
 +{{1}\over{r}}{{\partial\,\omega}\over{\partial\,r}}
               \Biggr).
\label{eq==6}
\end{equation}
 A general solution of this equation has the following view
\begin{widetext}
\begin{equation}
\label{eq==7}
 \omega(r,t)={{\mit\Gamma}\over{4\Sigma(\nu,t,\sigma)}}
  \exp
  \matrix{
  \left\{\displaystyle
      -{{r^2}\over{4\Sigma(\nu,t,\sigma)}}  
      \right\}},
\end{equation}
\begin{equation}
\label{eq==8}
\hspace{-8pt}
 {\mathit {v}}(r,t)
= {{1}\over{r}}\int\limits_{0}^{r}\omega(r',t)r'dr'
={{\mit\Gamma}\over{2 r}}
 \matrix{
             \left( 1 -
  \exp
  \left\{\displaystyle
      -{{r^2}\over{4\Sigma(\nu,t,\sigma)}}  
      \right\}
           \right).
          }
\end{equation}
\end{widetext}
 Here $\mit\Gamma$ is the integration constant having dimension [length$^2$/time]
 and the denominator ${\mit\Sigma}(\nu,t,\sigma)$ has a view
\begin{equation}
   {\mit\Sigma}(\nu,t,\sigma) =
   \int\limits_{0}^{t} \nu(\tau) d\tau + \sigma^{2}.
\label{eq==9}
\end{equation} 
 Here $\sigma$ is an arbitrary constant such that the denominator is always positive.
 The extra parameter~$\sigma$ comes from the Gaussian vortex cloud~\cite{KevlahanFarge1997} 
 with the circulation $\Gamma$ and the radius $\sigma$ initially existing.
 Thanks to this parameter, the vortex is protected from possible catastrophic finalization.
\begin{figure}[htb!]
  \begin{picture}(200,300)(-30,10)
      \includegraphics[scale=0.45]{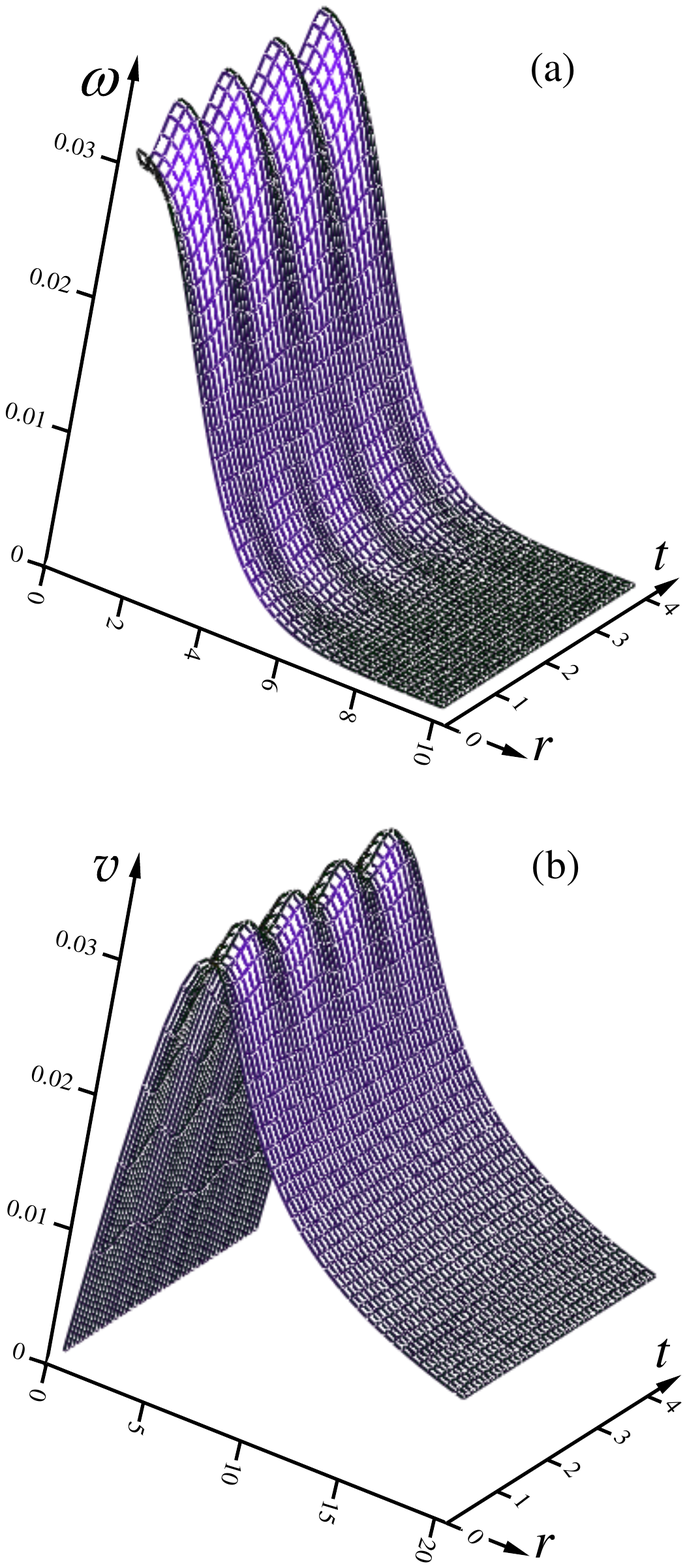}
  \end{picture}
  \caption{
  Vorticity $\omega(r,t)$ in (a) and velocity ${\mathit v}(r,t)$ in (b) as functions of $r$ and $t$
  for ${\mit\Gamma}=1$, $\nu=1$, $\Omega=2\pi$, and $n=16$. These parameters are conditional with aim to show qualitative picture of oscillations of the vortex in time.
   }
  \label{fig==2}
\end{figure}
 If the viscosity does not depend on the time, $\nu$=const and $\sigma=0$, this solution reduces to the Lamb-Oseen vortex solution~\cite{WuMaZhou2006}
which decays with the time. 
 
 On the other hand, if $\nu(\,t\,)$ is an alternating function of time, in average on the time it stays equal to zero, 
 then the vortex will live indefinitely long.
 Let, for the sake of simplicity, define this function as follows
\begin{equation}
  \nu(\,t\,) = \nu\cos(\Omega\,t )
=\nu{{e^{{\bf i}\Omega\,t}+e^{-{\bf i}\Omega\,t}}\over{2}}
\label{eq==10}
\end{equation}
 where $\Omega$ is an oscillation frequency.
Then the solution~(\ref{eq==7})-(\ref{eq==8}) takes the form:
\begin{eqnarray}
\nonumber
\hspace{-32pt}
   \omega(r,t) &=&
{{\mit\Gamma}\over{4 ({{\nu}/{\Omega}})(\sin(\Omega\,t)  +n ) }} \\
&&\times \exp\Biggl\{
  - {{r^2}\over{4 ({{\nu}/{\Omega}})(\sin(\Omega\,t) + n  ) }}
  \Biggr\},
\label{eq==11}
\end{eqnarray}
\begin{equation}
 {{\mathit v}}(r,\,t) = 
 {{\mit\Gamma}\over{2 r}}\Biggl(\!1 - 
 \exp{\Biggl\{
   - {{r^2}\over{4 ({{\nu}/{\Omega}})(\sin(\Omega\,t) + n ) }}
  \Biggr\}}\!
  \Biggr).
\label{eq==12}
\end{equation}
 Here an arbitrary dimensionless constant $n$ should be greater then 1, 
and $({{\nu}/{\Omega}})\cdot n=\sigma^{2}$ has dimension [length$^2$].
These solutions as functions of $r$ and $t$ are shown in Figs.~\ref{fig==2}(a) and~\ref{fig==2}(b), respectively.
One can see, that these functions undergo oscillations on the course of time.
 They do not attenuate, by demonstrating the oscillating mode.

\begin{figure}
  \begin{picture}(200,60)(0,15)
      \includegraphics[scale=0.5]{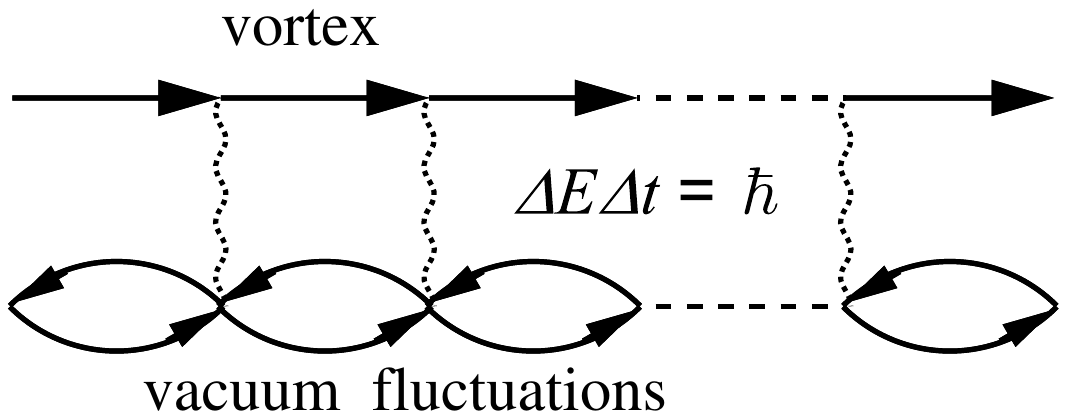}
  \end{picture}
  \caption{
  Periodic energy exchange between the vortex and vacuum fluctuations  }
  \label{fig==3}
\end{figure}

The alternating function of time, $\nu(t)$, representing the kinematic viscosity describes exchange of the stored energy of the vortex with the zero-point vacuum fluctuations, Fig.~\ref{fig==3}.
 In the average, the viscosity of the fluid medium is equal to zero. It means that this medium is superfluid. Such an example is well known: at transition of helium to the superfluid phase~\cite{Volovik2003}, coherent Cooper pairs of electrons are formed due to the exchanging phonons, which play a role of fluctuations of a background medium. 

\begin{figure}
  \begin{picture}(200,170)(0,5)
      \includegraphics[scale=0.5]{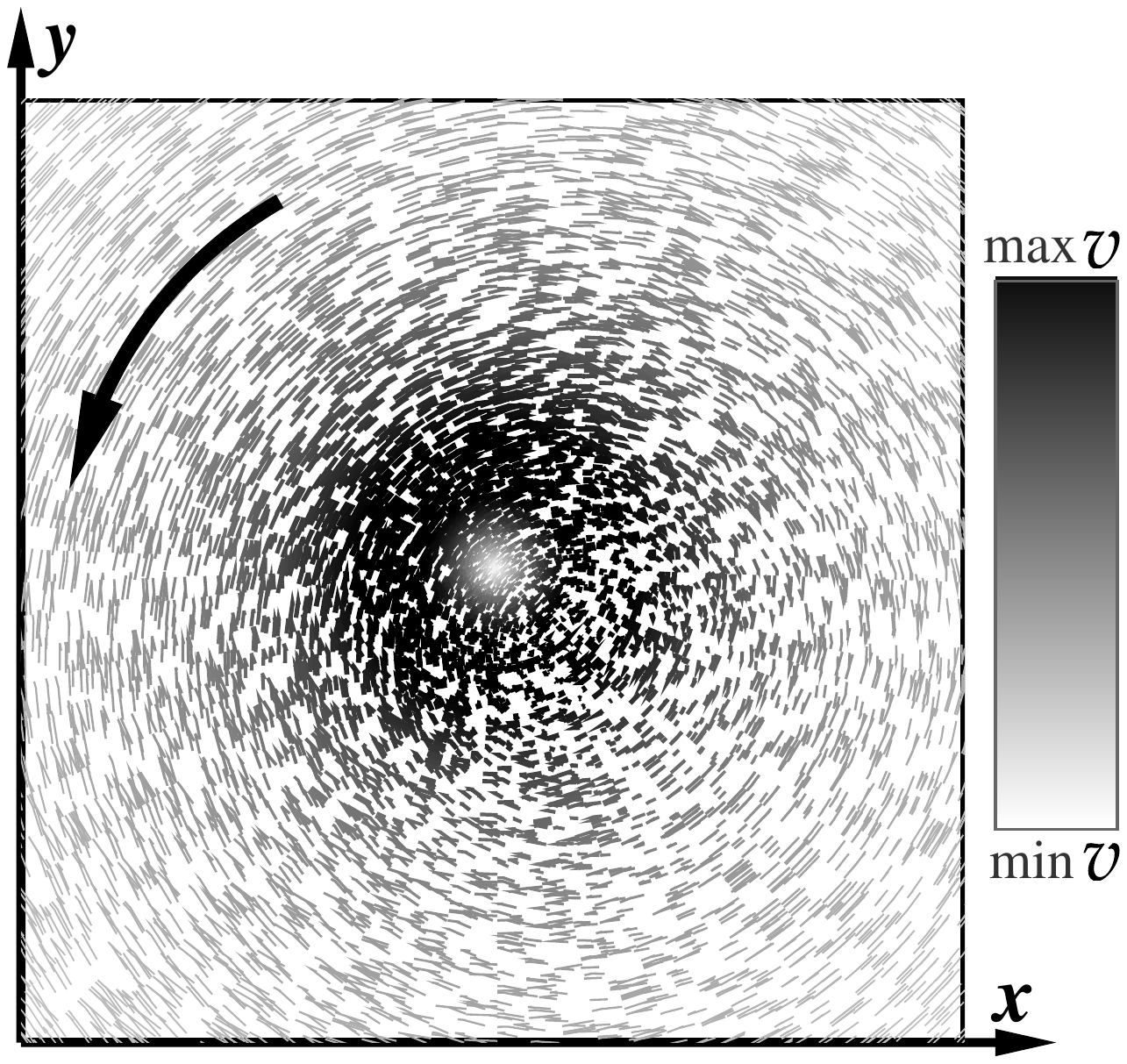}
  \end{picture}
  \caption{
  Cross-section of the vortex tube in the plane $(x, y)$. Values of the speed $\mathit v$ are shown in gray ranging from light gray
 ($\min\mathit v$) to dark gray ($\max\mathit v$).  Density of the pixels represents magnitude of the vorticity.
 Core of the vortex $\omega$ is well visible in the center.
  }
  \label{fig==4}
\end{figure}
  Qualitative view of the vortex tube in its cross-section is shown in Fig.~\ref{fig==4}. Values of the speed $\mathit v$ are shown by gray color ranging from light gray (minimal speeds) to dark gray (maximal speeds).  In the center of the vortex, the vortex core (so-called "eye of the hurricane"
   (see in Wikipedia: `Tropical cyclone'))
    is well viewed. Here it looks as a small light gray disk, where the speeds have small values. In the very center of the disk, in particular, the speed vanishes. This is in stark contrast to conditions in the region where moving the medium occurs most fast (in Fig.~\ref{fig==4} it looks as a dark gray annular region enclosing the light gray inner area). 
This transition region is named the wall of the core.

 To evaluate the radius of the core we need to equate to zero
the first derivative by $r$ of Eq.~(\ref{eq==12}) and to find its roots. Among two roots
 we find the root giving the interested radius~\cite{Sbitnev2015b}
\begin{equation}
\label{eq==13}
 r_{\rm core} \approx 2\sqrt{a_{0}( n+\sin(\Omega\,t))}\sqrt{ {{\nu}\over{\Omega}} }
\end{equation}
 Here $a_{0}\approx 1.2564$ is a root of the equation $\ln(2a_{0}+1)-a_{0}=0$.
One can see that the radius is an oscillating function of time.  As $\Omega$ is increased, the core radius decreases. 
However it grows with increasing $n$. At that, with $n\gg 1$ amplitude of the oscillations becomes negligibly small.

The trembling of the vortex radius due to exchange of the stored energy with the zero-point vacuum fluctuations creates, perhaps, additional variations of the pressure in the vacuum leading to the interaction of the vortices in a flow. Observe that pair of interacting vortices having opposite chiralities, forms a quasi-object, like the Cooper pairs  
which exist due to the interaction by exchanging phonons~\cite{Volovik2003}.

\section{\label{sec3}Vortex rings and vortex balls}

\begin{figure}
  \begin{picture}(200,110)(0,5)
      \includegraphics[scale=0.33]{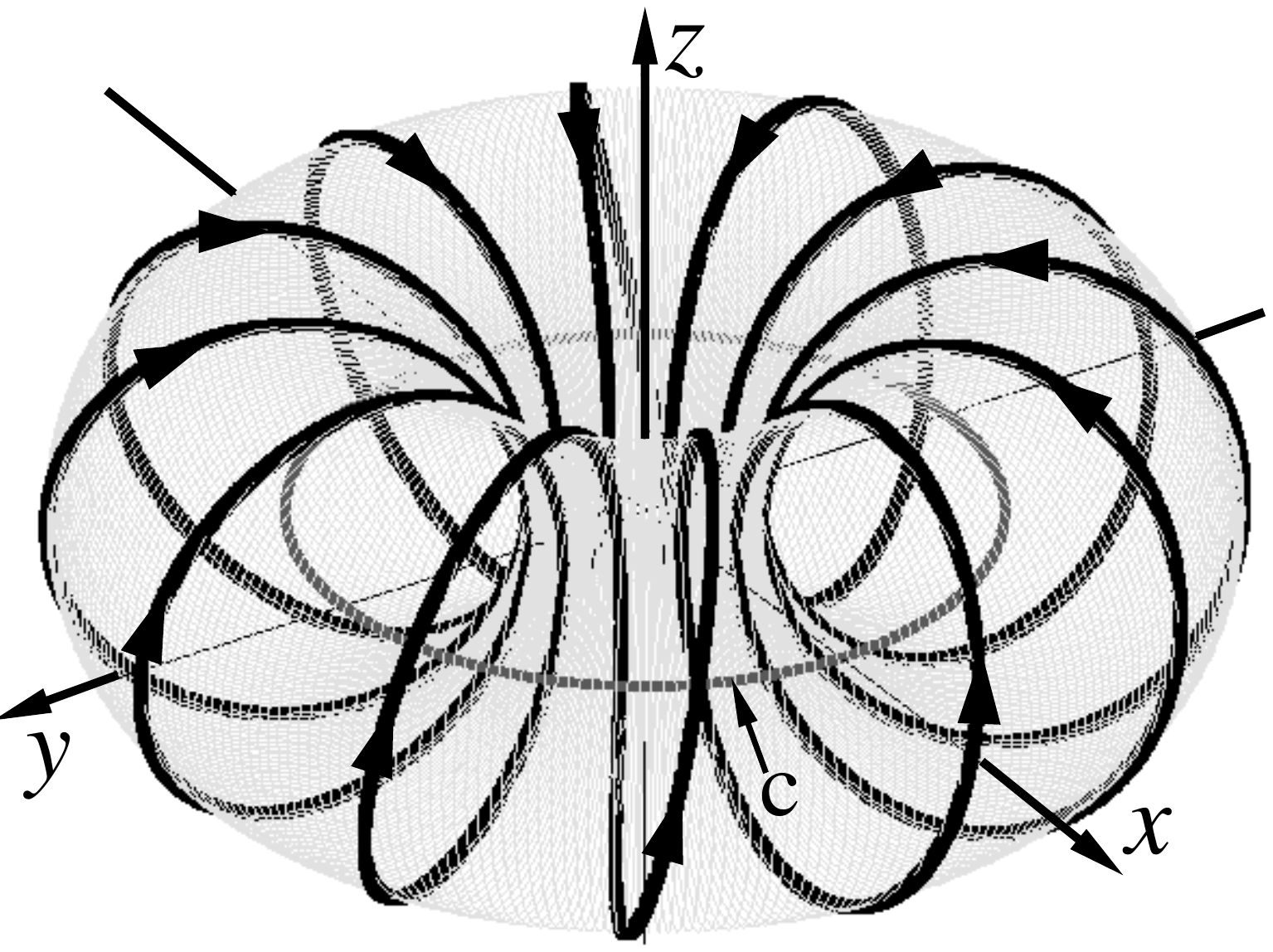}
  \end{picture}
  \caption{
  Helicoidal vortex ring is drawn on the wall of the vortex core: $a_{0}=2$, $b_{1}=3$, $\omega_{\,0}=12\,\omega_{1}$, $\phi_{\,0}=\phi_{1}=0$.  The wall, for the sake of visualization, is colored in light-gray. }
  \label{fig==5}
\end{figure}

 If we roll up the vortex tube in a ring and glue together its opposite ends
 we obtain a helicoidal vortex ring~\cite{Sonin2012}.
 Position of points on the vortex ring in the Cartesian coordinate system is given
 by the following set of equations
\begin{equation}
\label{eq==14}
\hspace{-14pt}
\left\{
\matrix{
  x=(b_{1}+a_{0}\cos(\omega_{\,0}t+\phi_{\,0}))\cos(\omega_{1}t+\phi_{1}),\cr
  y=(b_{1}+a_{0}\cos(\omega_{\,0}t+\phi_{\,0}))\sin(\omega_{1}t+\phi_{1}),\cr
  z= a_{0}\sin(\omega_{\,0}t+\phi_{\,0}).\hspace{90pt}\cr
       }
\right.
\end{equation}
The frequency $\omega_{\,0}$ is that of rotation about points lying on the center of the tube pointed by arrow c in Fig.~\ref{fig==5}. The frequency $\omega_{1}$ is that of rotation about the center of the torus localized in the origin $(x,y,z)$. Parameter $a_{0}$ is the radius of the tube. And $b_{1}$ is the radius of the torus - the distance from the origin to the circle pointed by arrow c in Fig.~\ref{fig==5}. 
The phases $\phi_{\,0}$ and $\phi_{1}$ can have different values between  0 to $2\pi$. 
The torus can be filled by helicoidal rings everywhere densely  by choosing different phases within this interval with a small increment.
 
 The  helicoidal vortex ring, with the parameters $a_{0}=2$, $b_{1}=3$, $\omega_{\,0}=12\,\omega_{1}$, and $\phi_{\,0}=\phi_{1}=~0$, is drawn by thick black spiral  in Fig~\ref{fig==5}.
It twists itself along the vortex wall dyed in light-gray.
The vorticity here has a maximal value on the center of the tube. 
The velocity of revolution  in the vicinity of this center is minimal, vanishes at the center and grows as a distance from the center increases. After reaching of some maximal value on the wall of the vortex core the velocity then  begins to decrease. 
\begin{figure*}[htb!]
  \begin{picture}(200,280)(60,5)
      \includegraphics[scale=0.56]{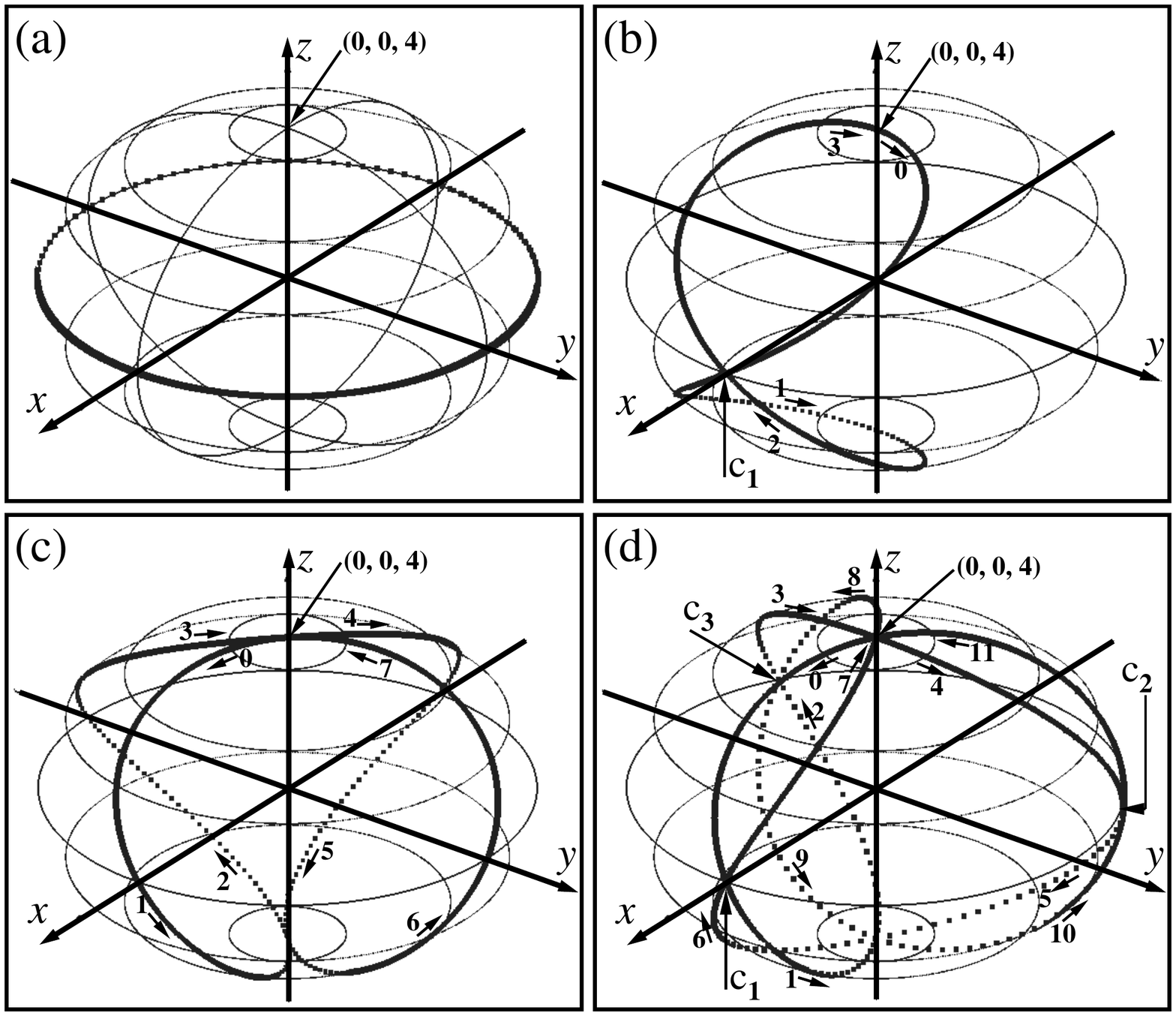}
  \end{picture}
  \caption{
Different modes of vortex rings having the radiuses $a_{0}=4$ and $b_{1}=0.01$, ($\phi_{\,0}=\phi_{1}=0$): (a) a degenerate loop, $\omega_{\,0}=0$.
It lies in the plane $(x, y)$ and never passes through the top and bottom poles; 
(b) single associated loop, $\omega_{1}=\omega_{\,0}$; (c) double-associated loop, $\omega_{1}=\omega_{\,0}/2$; (d) three times associated loop, $\omega_{1}=\omega_{\,0}/3$.
Arrows $0, 1, 2, \cdots$ show motions along the loops.
Arrows c$_1$, c$_2$, and c$_3$ printed in (b) and (d) point onto points of intersections of the loops with themselves.
Thin circles  emphasize the boundary of the vortex ball.
  }
  \label{fig==6}
\end{figure*}

\subsection{\label{subsec3.1}The vortex balls}

 Instead of the Hill's spherical vortex where all streamlines do not intersect~\cite{Tryggeson2007}, we shall consider vortex objects where intersections of the streamlines are allowed. These objects will be named the vortex balls.
( The name  "vortex ball" originates from the ball lighting - an astonishing electromagnetic bundle of energy emerging often during thunderstorms).

When the radius $b_{1}$ is greater than $a_{0}$, the vortex ring is the  torus,  Fig.~\ref{fig==5}. 
As soon as $b_{1}$ becomes smaller than $a_{0}$ the streamlines are penetrating into the inner part of the ring and begin to intersect with themselves.
Such an object in the limit $b_{1}\rightarrow 0$  converges to a sphere wherein the streamlines intersect each other on the sphere surface. It is the dipole surface.
Let now the radius $b_{1}$ in Eq.~(\ref{eq==14}) tends to zero. The helicoidal vortex ring in this case will transform into the vortex ring  enveloping a spherical ball  with the radius~$a_{0}$. Four different vortex rings for four different relationships of  $\omega_{\,0}$ vs. $\omega_{1}$ are drawn by thick curves in Fig.~\ref{fig==6}.

The arrows 0, 1, 2, etc.,   in Fig.~\ref{fig==6}  indicate paths of travel on the loops. 
Except of the degenerate loop, shown in Fig.~\ref{fig==6}(a), all loops begin from the pole $(x,y,z)=(0,0,4)$ and after a certain time are returned in the same top pole. At that, the return to the initial orientation of the arrow can occur after several passings through this pole with different orientations of the arrow in this pole. 
A simple return is shown in Fig.~\ref{fig==6}(b). 
Here the return to the initial orientation occurs after one revolution -  through the revolution on 360 degrees. 
In the case of the double-associated loop, Fig.~\ref{fig==6}(c), the return to the initial orientation  occurs after two revolutions. 
After the first revolution the arrow 0 changes its orientation to opposite. It is the arrow 3. After the second revolution the arrow comes to the position 7, which has the same initial orientation as the arrow 0. So, the full revolution is $2\cdot360=720$ degrees. 
Conditionally speaking, such an alternation of the orientations at each revolution on 360 degrees can be expressed by the alternation of opposite colors placed on the color wheel
 (Wikipedia: 'color wheel')
 say, green, magenta, green, magenta, etc.
Observe that this revolution shows a good agreement with the rotation of a 1/2--spin particle.
The return to the initial orientation of the arrow in the case of more complex loops requires multiple passings through the top pole with  different orientations of the arrow there. So, in the case of the triple-associated loop the return to the initial orientation occurs after three revolutions, Fig.~\ref{fig==6}(d). The full revolution  is equal to $3\cdot360=1080$ degrees.
 Changing orientation at each revolution on 360 degrees can be expressed, conditionally, in alternation of the primary color set: red, green, blue, red, green, blue, etc.
  In a general, the full return to the initial orientation in the top pole for the case of $n$-th associated loop can be achieved after the revolution on $ n\cdot$360 degrees.

\subsection{\label{subsec3.2}Rotation about the $z$-axis}

 Observe that in the case of odd $n$, the $n$-th associated loops intersect their own paths at points lying on the equatorial circle. In Figs.~\ref{fig==6}(b) and~\ref{fig==6}(d) these points are pointed by arrows c$_1$, c$_2$, c$_3$. It would seem that counter flows in these intersections should lead to destroying the vortex balls. However, sum of these flows will initiate rotation of the ball 
 about the $z$-axis. 
 
\begin{figure}[htb!]
  \begin{picture}(220,170)(-15,0)
      \includegraphics[scale=0.45]{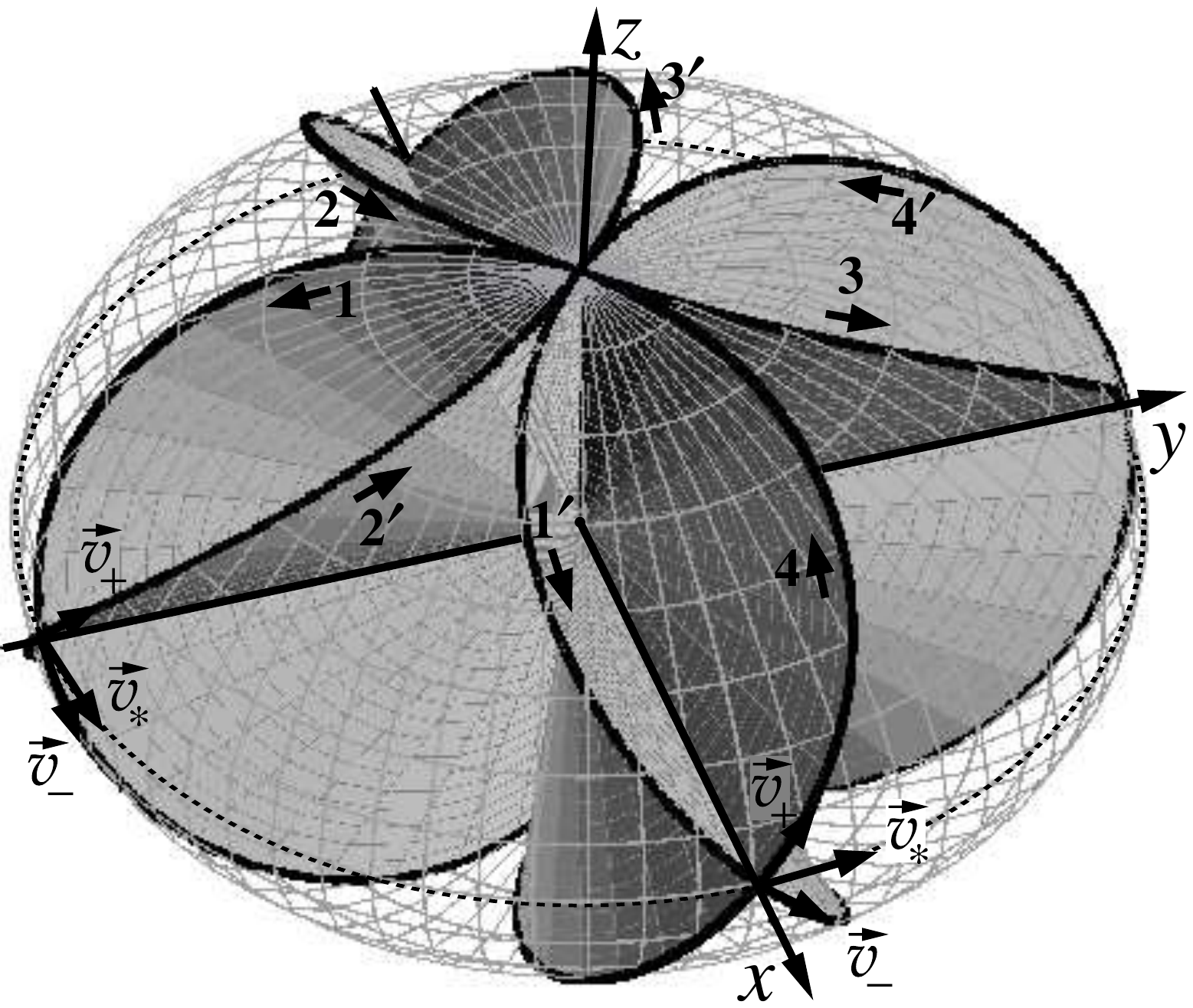}
  \end{picture}
  \caption{
 Two double-associated loops enveloping the sphere of unit radius, $a=1$, $b\approx 0$.
  }
  \label{fig==7}
\end{figure}

 In the case of even number $n$, these loops do not intersect themselves, as can be seen in Figs.~\ref{fig==6}(a) and~\ref{fig==6}(c). 
 However in the case of superposition of several $n$-th associated loops with different phase shifts $\phi_{\,0}$ and $\phi_{\,1}$ the intersection is possibly.
Such a case of the double-associated loops is shown in Fig.~\ref{fig==7}.
The two double-associated loops drawn by black thick curves envelop the sphere along its surface. 
These loops are shifted with respect to each other on the phase $\phi_{1} = \pi/2$, therefore their intersection takes place. 
The first loop (the phase is $\phi_{1} =0$) bypasses the path along the arrows 1 to 2 to 3 to 4 and further again to 1. 
The second loop (the phase is $\phi_{1} = \pi/2$) bypasses the path along the arrows 1' to 2' to 3' to 4' and further closes on~1'.
It should be noted that the paths are traversed through the poles always with changing orientation of the arrow in the opposite direction.
The arrow can take the same orientation only after it does that again, see Fig.~\ref{fig==6}(c) for the details.
\begin{figure}[htb!]
  \begin{picture}(220,140)(-30,0)
      \includegraphics[scale=0.3]{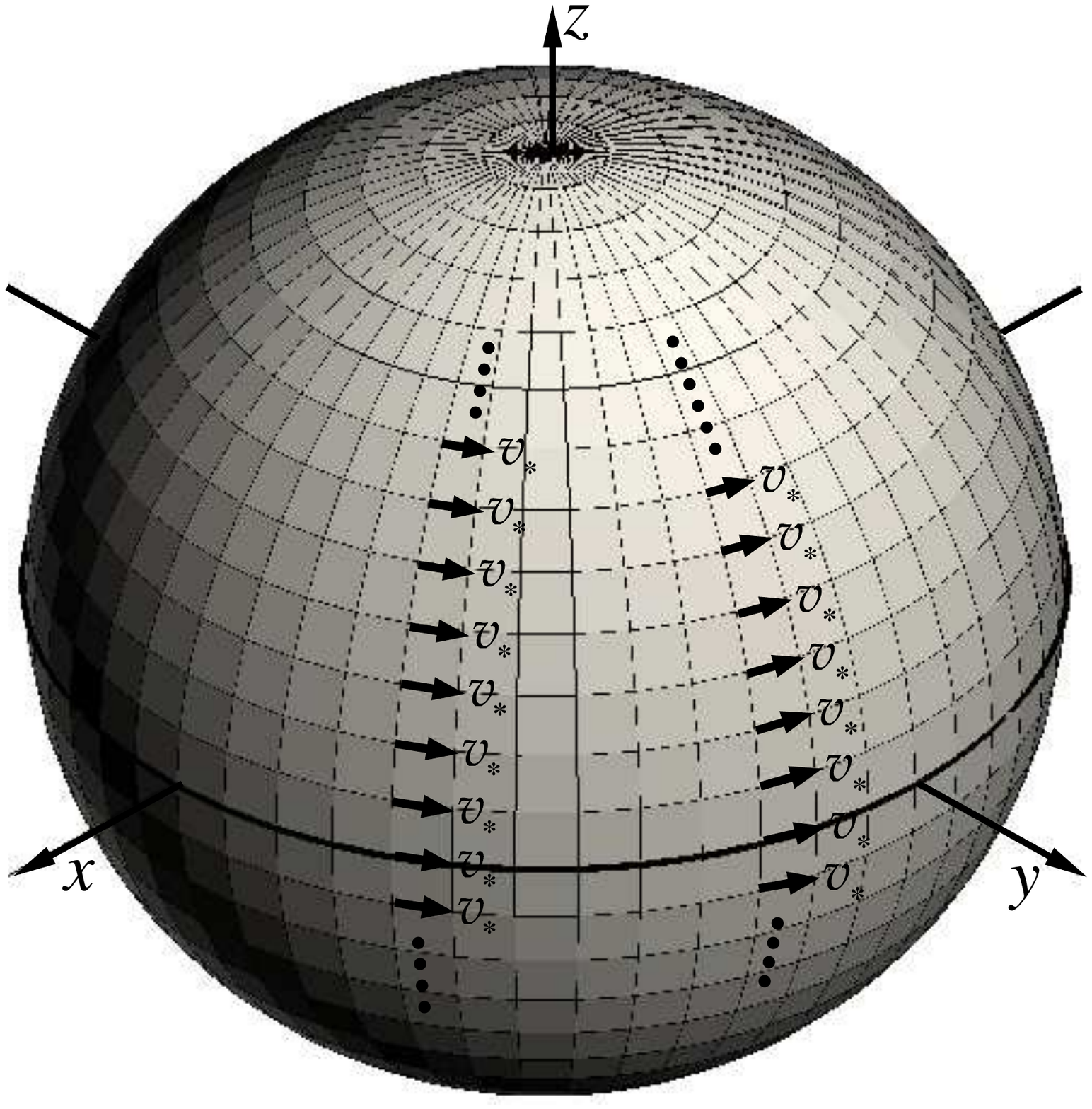}
  \end{picture}
  \caption{
 Qualitative presentation of the self-organizing flow on the wall of the vortex core for the case of the vortex ball:
 $a_{0}=1$, $b_{1}\approx 0$.
  }
  \label{fig==8}
\end{figure}

By choosing different phases $\phi_{\,0}$ and  $\phi_{\,1}$ from the interval $[0,2\pi)$ we can have different streams along both paths by initiating rotation of the ball about the $z$-axis with an orbital velocity  ${\vec{\mathit v}}_{*}$.
 It is equal to sum of two velocities of these counter streams, ${\vec{\mathit  v}}_{+}$ and ${\vec{\mathit v}}_{-}$ (Fig.~\ref{fig==7}).
These velocities are result of differentiation of Eq.~(\ref{eq==14}) by $t$.
Let us imagine that a small clot of the matter moves along the loop in the vicinity of the point $(x, y, z)$.
Then the velocity components, ${\mathit v}_{_{x}}$, ${\mathit v}_{_{y}}$, and ${\mathit v}_{_{z}}$, are as follows
\begin{widetext}
\begin{eqnarray}
   {\mathit v}_{_{x}} =
        &-& a_{0}\omega_{\,0}\sin(\omega_{\,0}t+\phi_{\,0})\cos(\omega_{1}t+\phi_{1})\cr
        &-& a_{0}\omega_{\,1}\cos(\omega_{\,0}t+\phi_{\,0})\sin(\omega_{1}t+\phi_{1})
        - b_{1}\omega_{1}\sin(\omega_{1}t+\phi_{1}),\cr
   {\mathit v}_{_{y}} =
        &-& a_{0}\omega_{\,0}\sin(\omega_{\,0}t+\phi_{\,0})\sin(\omega_{1}t+\phi_{1})\cr
       &+& a_{0}\omega_{\,1}\cos(\omega_{\,0}t+\phi_{\,0})\cos(\omega_{1}t+\phi_{1})
       +  b_{1}\omega_{1}\cos(\omega_{1}t+\phi_{1}),\cr
   {\mathit v}_{_{z}} =
       &~~&a_{0}\omega_{\,0}\cos(\omega_{\,0}t+\phi_{\,0}),
\label{eq==15}
\end{eqnarray}
\end{widetext}
Intersection points lying on the equatorial circle are points where superpositions of the velocities ${\vec{\mathit v}}_{+}$ and ${\vec{\mathit v}}_{-}$ give the summary velocity ${\vec{\mathit v}}_{*}$ oriented along the equatorial circle,
 that is,  ${\vec{\mathit v}}_{*}={\vec{\mathit v}}_{+}+{\vec{\mathit v}}_{-}=(2{\mathit v}_{_{x}}, 2{\mathit v}_{_{y}},0)$.
 The first velocity, ${\vec{\mathit v}}_{+}$, is calculated with the given phases $\phi_{\,0}=0$, $\phi_{\,1}=0$,
 and  the  second velocity, ${\vec{\mathit v}}_{-}$,  is calculated with  the phases $\phi_{\,0}=0$, $\phi_{\,1}=\pi/2$.
Infinitesimal shifts along these streamlines have the following values:
\begin{equation}
 \delta{\vec{\mathit r}}_{\pm} = {\vec{\mathit v}}_{\pm}{\delta t}.
\label{eq==16}
\end{equation}
 Here ${\delta t}$ represents the infinitesimal time shift.
The new position 
\begin{equation}
 {\vec{\mathit r}}_{*} = {\vec{\mathit r}}_{0} + \delta{\vec{\mathit r}}_{+} + \delta{\vec{\mathit r}}_{-} =
 {\vec{\mathit r}}_{0} + {\vec{\mathit v}}_{*}{\delta t}
\label{eq==17}
\end{equation}
lies on the same circle with the initial point ${\vec r}_{0}$. The circle lies in the plane perpendicular to the $z$-axis --  the motion occurs around the  $z$-axis.

By choosing the loops with other phases $\phi_{0}$ and $\phi_{1}$ one can cover the surface of the sphere by the intersection points  everywhere densely. In these points sum of the velocities ${\vec{\mathit v}}_{+}$ and ${\vec{\mathit v}}_{-}$ gives the summary velocity ${\vec{\mathit v}}_{*}$  lying in the plane $(x,y)$. All these velocities are oriented in the same direction about the $z$-axis. So that, the resulting rotation will occur about the $z$-axis, as shown in Fig.~\ref{fig==8}.
Note that this self-organizing flow rotates about the $z$-axis  repeating after each revolution on 720 degrees. 
Figuratively speaking, the vortex ball changes its color at each revolution on 360 degrees (say, green to magenta, magenta to green, etc.). 
We may imagine the 1/2 spin with installed to its tip a flag which rotates on the surface of 3D sphere in such a manner. The flag, at each revolution on 360 degrees, changes its orientation on opposite. It is picture of the spin rotation in the quaternionic representation~\cite{AgamalyanEtAl1988}.

\section{\label{sec4}Object spin-1/2  and the anomalous magnetic moment of electron}

Let us look on these transformations with perspective of the  group SU(2).
First we define the angular velocity, ${\vec{\Omega}}$ as the rate of change of angular displacement of the flag (harpoon-arrow) on the sphere with the radius $a_{0}$:
\begin{equation}
  {\vec{\Omega}} ={{{\vec{\mathit r}}\times{\vec{\mathit v}}_{_R}}\over{~|{\mathit r}|^2}}\Biggr|_{r = a_{0}}
\label{eq==18}
\end{equation}
Now we rewrite Eq.~(\ref{eq==15}) in assumption $b_{1}\rightarrow 0$, $\omega_{1}=\omega_{\,0}/2$, and let the phases, $\phi_{0}$ and $\phi_{1}$, be absent  (hereinafter, for the sake of simplicity, we shall omit subscript 0). We have:
\begin{widetext}
\begin{eqnarray}
\nonumber
 {\Omega}_{+} &=& {\Omega}_{x} + {\bf i}{\Omega}_{y} 
 = - {{1}\over{2}} \omega \exp\biggl\{{\bf i}{{\omega}\over{2}} t\biggr\}
 \Bigl(
          2\sin(\omega t)
         -{\bf i}\cos(\omega t )
 \Bigr),\\
\nonumber \\
\nonumber
 {\Omega}_{-} &=& {\Omega}_{x} - {\bf i}{\Omega}_{y}  
 =  -{{1}\over{2}} \omega \exp\biggl\{-{\bf i}{{\omega}\over{2}} t\biggr\}
 \Bigl(
            2\sin( \omega t)
       +{\bf i}\cos(\omega t)
 \Bigr),\\
\nonumber \\
 {\Omega}_{z} &=&  \omega\,
                         \cos(\omega t).
\label{eq==19}
\end{eqnarray}
\end{widetext}
 Here $\Omega_{x}={\mathit v}_{x}/a_{0}$, $\Omega_{y}={\mathit v}_{y}/a_{0}$,
 $\Omega_{z}={\mathit v}_{z}/a_{0}$.

Now we may write the equation describing behavior of the spin-1/2 under influence of a field $(\hbar\Omega_{x}, \hbar\Omega_{y}, \hbar\Omega_{z})$~\cite{Sbitnev2008}
\begin{equation}
\hspace{-6pt}
 {\bf i}\hbar {{\partial}\over{\partial\,t}} \hspace{-3pt}
 \left(
 \begin{array} {c}
 \psi_{\uparrow} \\
 \psi_{\downarrow}
 \end{array}
 \right)  =
 \left(
 \begin{array} {c c}
 \theta_{z} & \theta_{x} - {\bf i}\theta_{y} \\
 \theta_{x} + {\bf i}\theta_{y} & -\theta_{z}
 \end{array}
 \right) 
\hspace{-6pt}
 \left(
 \begin{array} {c}
 \psi_{\uparrow} \\
 \psi_{\downarrow}
 \end{array}
 \right).
\label{eq==20}
\end{equation}
 Here the energy coefficients $(\theta_{x},\theta_{y},\theta_{z})$ stem from 
$(\Omega_{x},\Omega_{y},\Omega_{z})$ by multiplying the latter by $\hbar$. More definitely $\hbar\omega$ represents the energy level of the rotation about $z$-axis.

We may continue this consideration by introducing the magnetic field~${\vec{\mathit B}}$~\cite{Martins2012} that acts on the spin-1/2 particle in the following manner $\vec\theta = \mu_{e}(\vec\sigma, {\vec{\mathit B}})$.
Here $\vec\sigma$ is the 3-component operator consisting of $2\times2$ Pauli matrices $\sigma_{x}$,  $\sigma_{y}$, and $\sigma_{z}$. 
Let $\mu_{e}$ be in the first approximation, the Bohr magneton
\begin{equation}
 \mu_{_{B}} = {{e\hbar}\over{2m_{e}}} \approx -9.27401452557\times10^{-24}~~{\rm J\cdot\!T}^{-1},
\label{eq==21}
\end{equation}
 which gives the magnetic moment of the Dirac electron.
 Here $e$ is the electron charge (it is negative) and $m_{e}$ is its mass.
 System of units adopted here is SI.
\begin{figure}
  \begin{picture}(200,70)(-10,5)
      \includegraphics[scale=0.4]{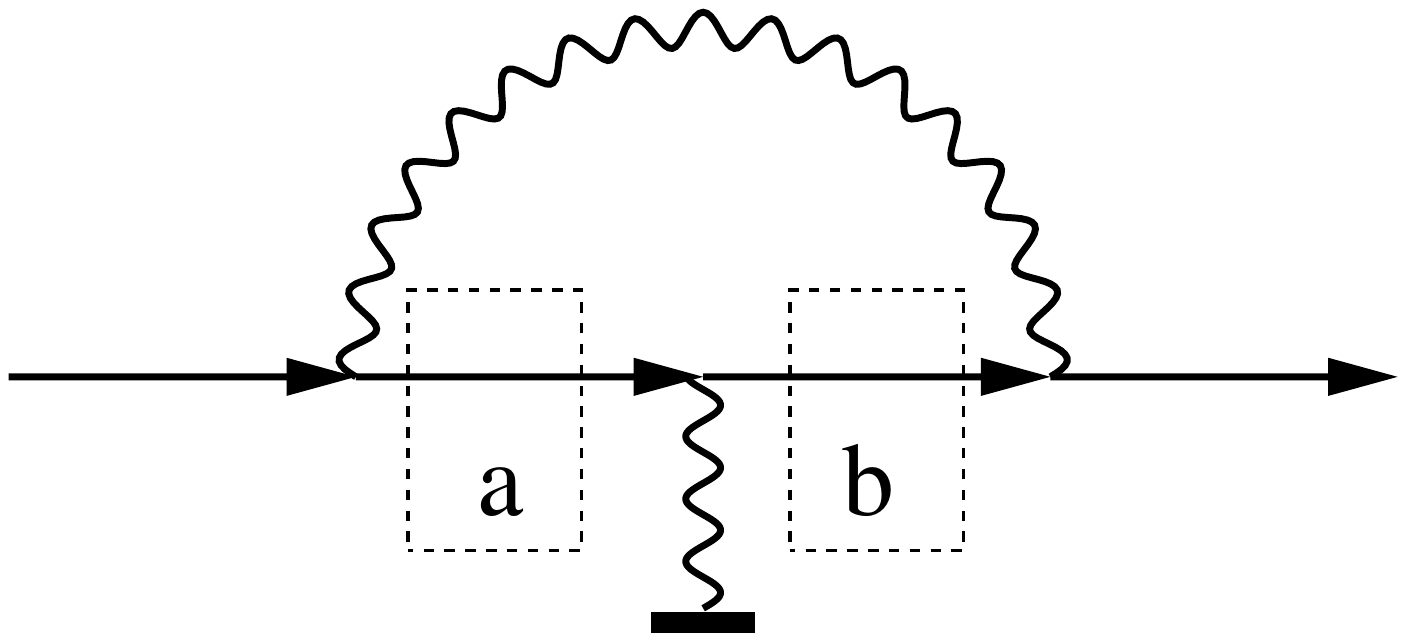}
  \end{picture}
  \caption{
The Feynman diagram showing the second-order contribution
to the anomalous magnetic moment. Rectangles a and b drawn by dotted lines  highlight areas where the similar second-order diagrams can be added. Thick line drawn below the wavy line marks the zero-point vacuum background 
(as a symbol of the ground in electronic circuit diagrams).
  }
  \label{fig==09}
\end{figure}
\begin{figure}
  \begin{picture}(200,130)(0,5)
      \includegraphics[scale=0.4]{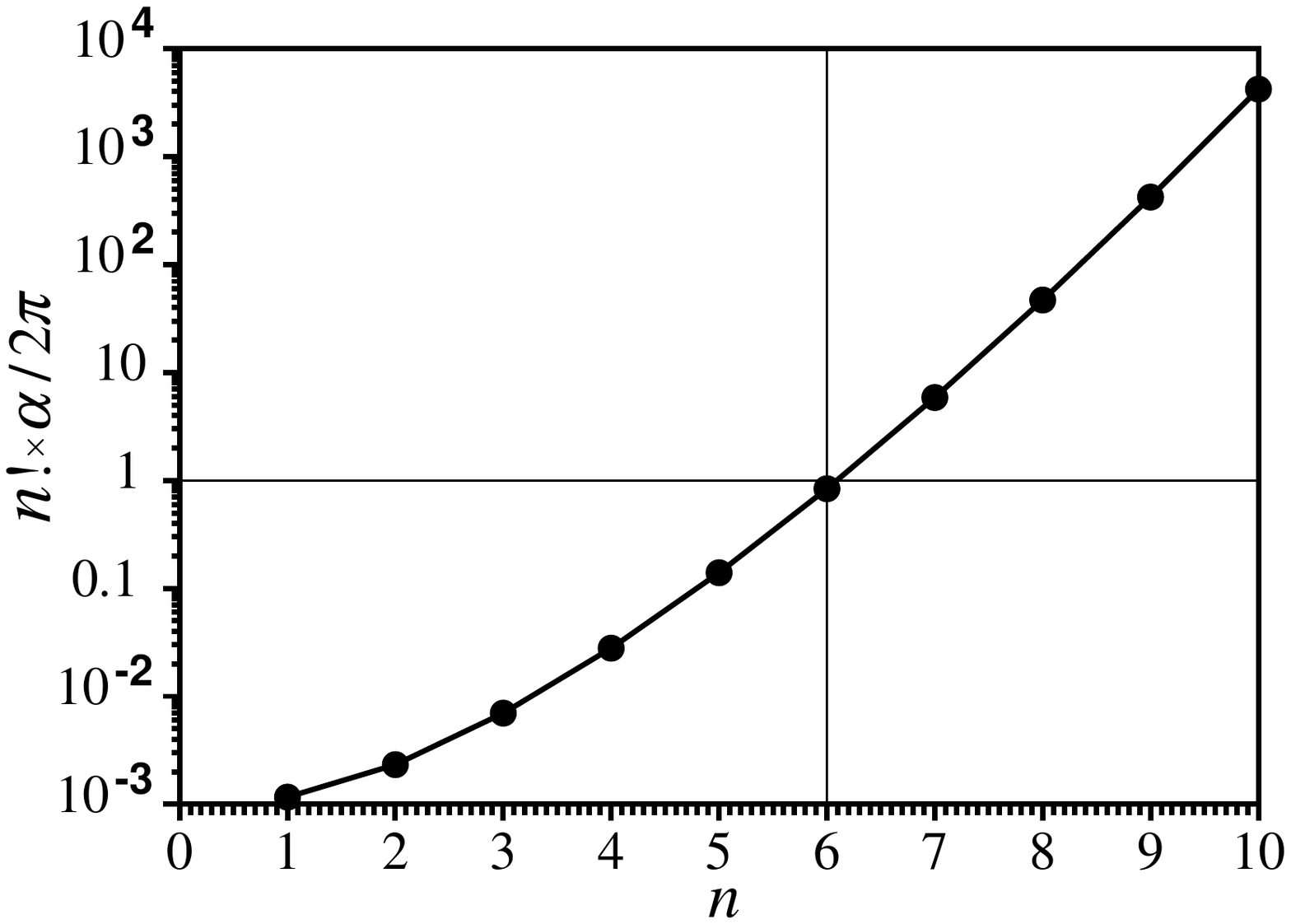}
  \end{picture}
  \caption{
  The term $n!\alpha/2\pi$ versus $n$.
  }
  \label{fig==10}
\end{figure}

Observe, however, that the electron has the anomalous magnetic moment that slightly differs from the above value.
This difference is induced by interaction of the charged particle with
the zero-point oscillations of the electromagnetic field~\cite{Mandache2013}.
This interaction occurs due to exchange of the particle energy with the  zero-point vacuum fluctuations.
As for our case, it is manifested in that the vortex draws into a whirl vacuum fluctuations which can be positioned far away of the vortex core.

Schwinger was first to compute the anomalous magnetic moment in 1947~\cite{Schwinger1948}
\begin{eqnarray}
\nonumber
\hspace{-24pt}
  \mu_{e} &=& \mu_{_{B}}(1 + \alpha/2\pi) \\
   &\approx&  -9.28478137856\times10^{-24}~~{\rm J\cdot\!T}^{-1}.
\label{eq==22}
\end{eqnarray}
He has applied the one-loop correction for calculation of the anomalous magnetic moment as shown in  Fig~\ref{fig==09}.
 Here $\alpha = e^2/(4\pi\epsilon_{0}\hbar\,c)$ is the fine structure constant, $\epsilon_{0}$ is the vacuum permittivity,
 and $c$ it the light speed. 
In fact, Schwinger has shown, that the fine structure constant (more definitely, the term $\alpha/2\pi$) plays an important role at calculating this moment.
 
 Sommerfield in 1958 has calculated a more exact value of the anomalous magnetic moment~\cite{Sommerfield1958},
 which has the following representation
\begin{eqnarray}
\nonumber
\hspace{-24pt}
 \mu_{e} &\approx&  \mu_{_{B}}(1 + \alpha/2\pi - 1.312(\alpha/2\pi)^2) \\
&\approx&  -9.284764966\times10^{-24}~~{\rm J\cdot\!T}^{-1}.
\label{eq==23}
\end{eqnarray}
 For comparison,  the most accurate value for the electron magnetic moment is as follows
 $\mu_{e}= ( -9.28476377\pm 0.000~000~23)\times10^{-24}~~{\rm J\cdot\!T}^{-1}$.
(This  value is given in 
 Wikipedia: `Electron magnetic moment'). 

One may guess that the term $\alpha/2\pi$ shows itself in fractal cascades of the second-order diagrams.
 We may highlight areas in the second-order diagram, Fig.~\ref{fig==09}, where the similar  diagrams can be realized. They are drawn by dotted rectangles. 
 A contribution of these diagrams falls with more deep their location.
 However amount of these diagrams can grow as~$n!$.
 Such a fractal structure can be reproduced through the continued fraction:
 \begin{widetext}
\begin{equation}
\mu_{e} =  \mu_{_{B}}
 {                            1      \over\displaystyle 1 -
 {\strut         \alpha/2\pi \over\displaystyle 1 +
 {\strut   2!\,\alpha/2\pi \over\displaystyle 1 -
 {\strut   4!\,\alpha/2\pi \over\displaystyle 1 -
 {\strut   6!\,\alpha/2\pi \over\displaystyle 1 +
 {\strut   8!\,\alpha/2\pi \over\displaystyle 1 -
 {\strut 10!\,\alpha/2\pi \over\displaystyle 1 -
 {\strut 12!\,\alpha/2\pi \over\displaystyle 1 +
 {\strut 14!\,\alpha/2\pi \over\displaystyle 1 -
 {\strut 16!\,\alpha/2\pi \over\displaystyle 1 -
 {\strut 18!\,\alpha/2\pi \over\displaystyle 1 + \cdots
 }}}}}}}}}}}.
\label{eq==24}
\end{equation}
\end{widetext}
Note that signs in this  fraction alternate as follows $-,-,+,-,-,+,\cdots$, the Bohr magneton 
 has  negative sign,
 see Eq.~(\ref{eq==21}).
The calculation of this fraction up to the contribution of the member $1- 18!\,\alpha/2\pi$ gives the following result $\mu_{e}\approx  -9.28476339\times10^{-24}~~{\rm J\cdot\!T}^{-1}$.
One may guess, however, that not all terms in this continued fraction can give correct contributions. Sooner or later the term 
$n!\alpha/2\pi$ will exceed unit. Fig.~\ref{fig==10} shows growth of this term as $n$ increases.
 We see that at $n=6$ the value of $n! \alpha/2\pi$ reaches almost 1 and further it grows catastrophically. So that $n=6$ is the limiting number, higher of which calculated values of the continued fraction~(\ref{eq==24}) lose sense.
 A value of the continued fraction calculated up to the limiting number $n=6$ 
  is as follows $\mu_{e}\approx  -9.28476377\times10^{-24}~~{\rm J\cdot\!T}^{-1}$.
 
\section{\label{sec5}Conclusion}

The modified Navier-Stokes equation with the two slightly changed terms~\cite{Sbitnev20xx} - the internal pressure and the viscosity of the superfluid medium, describes a wide spectrum of motions in this medium. 
The medium represents physical vacuum, consisting of enormous amount of virtual particle-antiparticle pairs. 
These particle-antiparticle pairs constitute the Bose-Einstein condensate which fill all the space.

The internal pressure divided by the density distribution of the particle-antiparticle pairs is the quantum potential. 
So, the spooky action at the distance is achieved due to the osmotic pressure arising in this superfluid medium at its motion.
That fact that the physical vacuum is the superfluid medium, it follows that its viscosity vanishes. 
We guess that time-averaged the viscosity  is zero. But its variance is not zero. 
It means that there is a regular exchange of energy of a particle with the zero-point vacuum fluctuations.

A particle in question can be a vortex being created from the vacuum medium. More definitely, 
the particle is the helicoidal vortex ring.
(One can guess that the helicoidal vortex rings are similar to strings~\cite{YauNadis2010} after reducing them onto the 3D-brane, to our three-dimensional space).
 Topologically the vortex ring can be transformed further to the vortex ball (in some sense it is the vortex bubble) having two poles - up and down. Flows on such an object can pose complicated patterns.
It is important to emphasize, that due to  viscosity fluctuations near zero, 
exchange between the vortex and the zero-point vacuum fluctuations takes place. 
Because of such an exchange, the vortex ball trembles and lives as long as possible. Such a trembling can provide interaction between different vortex objects due to the osmotic pressure arising in the medium. It is analogous to the exchange of phonons between the electrons forming the Cooper pairs in the superfluid helium~\cite{Volovik2003}.

The so-called vortex wall represents a spherical surface of the vortex ball where the vortex flows reach a maximal orbital speed.
Below this surface, the orbital speed decreases monotonically up to zero  when moving to the center of the ball. 
As we move from the ball onto infinity the orbital speed decreases monotonically up to zero also. 
The flows on the vortex walls can pose complicated patterns. An interesting case is when the pattern turns inside out after rotation on 360 degrees and returns to the initial position after repeated rotation on 360 degrees. Conditionally, one may imagine as periodic change of the wall color after each revolution on 360 degrees, say from  green to magenta, from magenta to green, etc.

In conclusion we may outline the following resume: the modified Navier-Stokes equation in pair with the continuity equation describes flows of the special superfluid medium~\cite{Sbitnev20xx} in which two flows are identified - irrotational and solenoidal flows. 
The first flow is conditioned by existence of a scalar field. This field represented by the action $S$ determines mobility of a particle through gradient of this function. Whereas coordinates of the particle can be measured accurate to its probabilistic position, through evaluations of the probability density distribution. These two quantities are represented in the complex-valued wave function which is the solution of the Schr{\"o}dinger equation. 
The second flow is the solution of the vorticity equation stemming from the Navier-Stokes equation. 
This equation can give a set of vortex solutions that evolve against the background of the scalar field. 
Observe that the vortices are robust long-lived objects which tremble in time.

If we adopt that the vortices, like strings, represent particles, then in the light of the results, outlined in the previous article~\cite{Sbitnev20xx} motion of these particles is in good agreement with 
de Broglie-Bohm theory~\cite{BensenyEtAl2014, Bush2015a, deBroglie1987}.
According to this theory the wave function is a pilot wave that guides the vortex 
along a most optimal path~\cite{FeynmanHibbs1965} from a  source to a detector.
The communication can be as follows. The vortex trembles and due to this perturbs virtual particle-antiparticle pairs, inhabiting the superfluid vacuum medium. It leads to forming the constructive and destructive interference
induced by a synergistic effect of these pairs. 
Through the wave function the vortex obtains back information about its state in the space through which it moves.

\begin{acknowledgements}
The author thanks Denise Puglia, Mike Cavedon, and Pat Noland for useful and valuable remarks and offers.
The author thanks the FOOP's reviewers for the constructive critique and proposals
\end{acknowledgements}

\bibliographystyle{spmpsci}      

%
%

\end{document}